\documentclass[twocolumn,nofootinbib,aps,pra,groupedaddress]{revtex4}
\usepackage{epsfig,amssymb,amsmath,mathtools}

\def\labell#1{\label{#1}}
\def\>{\rangle}\def\<{\langle}
\def\togli#1{}

\begin{document}
%\fbox{{\scriptsize  Preliminary. \today.}}

\title{ A quantum algorithm for estimating the determinant}
 \author{Vittorio Giovannetti,$^1$ Seth
  Lloyd$^{2}$, Lorenzo Maccone$^3$}\affiliation{$^{1}$NEST-INFM \&
  Scuola Normale Superiore, Piazza dei Cavalieri 7, I-56126, Pisa,
  Italy.\\\vbox{$^{2}$Dept. of Mech. Eng., Massachusetts
    Institute of Technology,
    77  Mass. Av., Cambridge, MA 02139, USA.}\\
  \vbox{ $^3$ Dip.~Fisica ``A.  Volta'' \& INFN Sez.~Pavia,
    Universit\`a di Pavia, via Bassi 6, I-27100 Pavia, Italy.}}
\begin{abstract}
  We present a quantum algorithm for estimating the matrix determinant
  based on quantum spectral sampling.  The algorithm estimates the
  logarithm of the determinant of an $n \times n$ positive sparse
  matrix to an accuracy $\epsilon$ in time
  ${\cal O}(\log n/\epsilon^3)$, exponentially faster than previously
  existing classical or quantum algorithms that scale linearly in $n$.
  The quantum spectral sampling algorithm generalizes to estimating
  any quantity $\sum_j f(\lambda_j)$, where $\lambda_j$ are the matrix
  eigenvalues.  For example, the algorithm allows the efficient
  estimation of the partition function
  $Z(\beta) =\sum_j e^{-\beta E_j}$ of a Hamiltonian system with
  energy eigenvalues $E_j$, and of the entropy
  $ S =-\sum_j p_j \log p_j$ of a density matrix with eigenvalues
  $p_j$.
\end{abstract}
\maketitle

Quantum computers can provide exponential speedups over
classical computers for a variety of linear algebraic tasks, including
fast Fourier transforms, finding eigenvectors and eigenvalues, and
matrix inversion \cite{nielsenchuang,copper,setheig,hhl}.   Here we provide a quantum spectral
sampling algorithm, and apply it to 
a fundamental linear algebraic problem, that of estimating the determinant
of a large sparse positive matrix.    Estimating the value of the determinant of large matrices is a
  common procedure in linear algebra, with wide application in, e.g.,
  data analysis and machine learning.  Our algorithm 
is related to the power of a single qubit/qumode
algorithm, as well as to quantum matrix inversion. 
Starting with a single quantum
mode or a few quantum bits in a pure state, together with a quantum system
in a fully mixed state, 
quantum phase estimation allows us to sample uniformly from the eigenvalues
of Hermitian matrices.  
We use quantum phase estimation to sample from the
eigenvalues $\lambda_j$ of an $n\times n$ sparse, positive Hermitian matrix $A$, whose smallest eigenvalue is bounded
below by $\lambda_{\min}$, and whose largest eigenvalue is bounded
above by $\lambda_{\max}$, so that the condition number of the matrix
is bounded by $ \kappa = \lambda_{\max}/\lambda_{\min}$.  For
convenience, we rescale  ${A}$ so that $\lambda_{\max} = 1/2$ and
$\lambda_{\min} = 1/2\kappa$ (see App.~\ref{notation}).  We then use
Monte Carlo sampling to estimate its log-determinant, i.e. the quantity $\alpha:=\log \left( {\rm det} A\right)=\sum_{j=1}^n \log \lambda_j$. Assuming $A$ is encoded into a quantum random access memory \cite{qram,nielsenchuang,qram1,qram2,qram3,qram4,qram5,qram6} or that it is efficiently simulable \cite{hast,chuang}, we show that $\alpha$
can be estimated with a relative error $\epsilon$ in a number of computational steps
\begin{eqnarray}
\#_{\rm LgD}
\simeq {\cal O}\left(\frac{\kappa \;(\log \kappa)^2 \; s^2 }{\epsilon^3}  \log n\:\log^2(s^2/\epsilon)\right)
\labell{sec}\;,
\end{eqnarray} 
where 
 $s$ is the 
sparsity of the matrix $A$ (the maximum number of nonzero elements for each row or
column), while $\kappa$ bounds  its condition number  (the ratio between its largest and smallest eigenvalues).
The right-hand-side of~(\ref{sec}) formally corresponds to the time complexity $T^{\rm [ora]}_{\rm LgD}$ of 
the procedure,
as measured in an oracle model~\cite{berry3} that supplies the elements of $A$ in response to appropriate queries (see App.~\ref{s:app}).
Existing classical and quantum algorithms for
estimating $\alpha$ all scale
at least proportionally to $n$,
e.g.~\cite{Newref0,Newref1,Newref2,class1,class2,class3,fs0,fs1,fs2}, whence our
exponential-advantage claim. Our algorithm follows by merging the
quantum spectral sampling with quantum phase estimation to give a
Monte Carlo estimate of the determinant.
Moreover, the quantum spectral sampling algorithm also allows one to estimate
the sum of any function of the eigenvalues, $\sum_j f(\lambda_j)$,
thereby allowing estimates of the partition function for Hamiltonian
quantum systems and the entropy of a density matrix.  Because it only
requires a single purified quantum mode, or a small number of pure state
qubits, quantum spectral sampling is particularly parsimonious
in terms of the quantum resources used, and could be implemented
effectively on near term noisy quantum information processors without
quantum error correction. 

We first describe the quantum spectral sampling algorithm in terms of
estimating the determinant, and then generalize to sums of arbitrary
functions of eigenvalues below.  Suppose we are given a
suitable description of the matrix $A$, for example, an algorithm for
efficiently calculating its elements \cite{hast,chuang}, or a
description of its elements stored in a quantum random access memory
\cite{nielsenchuang,qram}, we can use standard techniques of quantum
simulation
\cite{sethscience,childs,dorittashma,berry,berry2,berry3,chuang,childs2,REF3}
to enact the time evolution operator $e^{-iAt}$ to accuracy $\eta$ in
a number of computational steps
$N_{\rm qs}={\cal O}(\tau \log n \log^2(\tau/\eta)/\log\log(\tau/\eta))$
\cite{berry3} (see also App.~\ref{s:app}), where $\tau=s^2t\|A\|_m$ with
$s$ the maximum number of non-zero entries in each row/column of ${\alpha}$
and $\|A\|_m$ an upper bound to the modulus of the entries of ${\alpha}$. In
our case we need ${t}\simeq 1$ and with the rescaling of ${\alpha}$, also
$\|A\|_m$ is of order one. So, neglecting the $\log\log$ term, we can
approximate $N_{\rm qs}={\cal O}(s^2\log n\log^2(s^2/\eta))$.

We use this ability to apply the quantum phase estimation algorithm to
an $n$-dimensional quantum system in the fully mixed state described
by density matrix $\openone/n$.  The quantum phase algorithm
\cite{ki,chiara} is a digital analogue of von Neumann's pointer
variable model of measurement \cite{von}.  Von Neumann's object in the
pointer variable model was to construct a physical interaction between
a system and a (continuous) pointer variable that correlates the
energy eigenstates of ${\alpha}$ with a value of the pointer variable that
contains an estimate of the corresponding energy eigenvalue: a
physical model of how to perform a von Neumann measurement.  In
particular, von Neumann considered the application of the Hamiltonian
$A\otimes P $, where $P$ is the momentum operator for the pointer
variable, to the initial state
$|\psi\rangle = \sum_j \psi_j |j\rangle \mathmbox{|x\!=\!0\rangle}$,
where $A |j\rangle = \lambda_j |j\rangle$.  Applying this Hamiltonian
for time $t$ results in the state
$\sum_j \psi_j |j\rangle \mathmbox{| x\! =\! \lambda_j t\rangle}$.
The power-of-a-single-quantum-mode algorithm encodes the pointer
variable in the continuous variable state of a quantized harmonic
oscillator, and then uses quantum simulation to enact the von Neumann
pointer variable Hamiltonian.

The quantum phase estimation algorithm \cite{ki,chiara} digitizes von
Neumann's model by substituting an $m$-qubit register with $2^m$
states for the continuous pointer variable, and employing the
discretized version of the momentum operator.  Applying Hamiltonian
quantum simulation for a time ${t}=O(1)$ [i.e., one iteration of the
phase estimation that requires $2^{m}$ applications of a controlled
$e^{-iAt}$ plus a final quantum Fourier transform], then yields the
state
$\sum_j \psi_j |j\rangle\mathmbox{ | x\! =\!  \tilde\lambda_j
  \rangle}$, where $\tilde\lambda_j$ is an estimate of $\lambda_j$
that is accurate to ${\cal O}(1/2^m)$, for a suitable choice of the initial
state of the discretized pointer variable~\cite{chiara}.  The quantum
phase estimation algorithm takes ${\cal O}(2^m)$ steps to
attain this accuracy.  Applying the quantum phase estimation algorithm
to the fully mixed state, and then measuring the pointer variable
register allows us to sample uniformly from the eigenvalues
$\lambda_j$ of $A$, and to determine each eigenvalue to a root mean square error (rmse) accuracy
 $\delta \lambda_j \propto {\cal O}(1/2^m)$.  The accuracy to which
the quantum phase estimation algorithm determines $\log \lambda_j$
then goes as
$\Delta(\log\lambda_j)=\delta\lambda_j|
\tfrac{\partial\log\lambda_j}{\partial\lambda_j}|=\delta \lambda_j/
\lambda_j \approx 1/(2^m \lambda_j) \leqslant \kappa/2^{m-1}$, where
$\kappa$ is the upper bound to the condition number as above, so that
the quantum phase estimation rmse on ${\alpha}$ is
$\Delta \alpha_{\rm qpe}=\sqrt{}\sum_j(\delta\lambda_j/\lambda_j)^2
\leqslant\sum_j|\delta\lambda_j/\lambda_j|\leqslant n\kappa/2^{m-1}$,
where we used the triangle inequality (we are only interested in the
order of magnitude).
We then use Monte Carlo sampling to estimate the log-determinant 
$\alpha$.
If we take $N_{\rm mc}$ samples $\{ \lambda_\ell \}$, then the estimate of
$\alpha$ is
$\alpha_{est}:=\tfrac n{N_{\rm mc}}\sum_{\ell=1}^{N_{\rm mc}}\log\lambda_\ell$,
which tends to ${\alpha}$ for large $N_{\rm mc}$, with a statistical rmse
$\Delta \alpha_{\rm mc}=n\Delta/\sqrt{N_{\rm mc}}$, where given
$\mu:= \frac{1}{n}  \sum_{j=1}^n \log \lambda_j = \alpha/n$ the arithmetic mean  of the
logarithm of the eigenvalues of $A$, 
 $\Delta^2:=\frac{1}{n}  \sum_{j=1}^n (\log \lambda_j - \mu)^2$ is the associated variance (see App.~\ref{notation}).
The total rmse for ${\alpha}$ can finally be bounded by the sum of $\Delta \alpha_{\rm mc}$ and  $\Delta \alpha_{\rm mc}$, i.e.
$\Delta \alpha_{\rm err}\leqslant\Delta \alpha_{\rm qpe}+\Delta \alpha_{\rm mc}$ (see
App.~\ref{s:appvit2}), so the relative error is
\begin{equation}
\frac{\Delta \alpha_{\rm err}}{|\alpha|}\leqslant\frac n{|\alpha|}
  \Big(\frac{2\kappa}{2^{m}}+\frac\Delta{\sqrt{N_{\rm mc}}}\Big) = \frac 1{|\mu|}
  \Big(\frac{2\kappa}{2^{m}}+\frac\Delta{\sqrt{N_{\rm mc}}}\Big)
\labell{pri}\;.
\end{equation}
By setting $N_{\rm mc}:=(\tfrac{2\Delta}{|\mu|\epsilon})^2$ and
$2^m=N_{\rm qpe}:=\tfrac{4\kappa}{|\mu|\epsilon}$, we ensure that
$\frac{\Delta \alpha_{\rm err}}{|\alpha|}$ is smaller than or 
 equal to $\epsilon$. With this choice 
 the total number of steps to attain accuracy $\epsilon$ goes
as 
\begin{equation}
\#_{\rm LgD} \simeq N_{\rm qs}N_{\rm qpe}N_{\rm mc}
={\cal O}\left(\frac{\kappa \; s^2 \; \Delta^2}{|\mu|^3\; \epsilon^3}  \log n\:\log^2(s^2/\eta)\right)
\labell{sec1}\;,
\end{equation} 
which reduces to Eq.~\eqref{sec} once we set $\eta\simeq\epsilon$ and
link the factor $\Delta^2/|\mu|^3$ to the condition number $\kappa$ via
the inequalities $|\mu|\geq \log 2$ (since
$\lambda_j\leqslant \tfrac12$) and $\Delta \leq (\log \kappa)/2$
(since the largest $\Delta$ is when half the eigenvalues are
$\lambda_{min}=\tfrac1{2\kappa}$ and half are
$\lambda_{max}=\tfrac12$), see App.~\ref{notation}.  We conclude by
noticing that when running the algorithm we do not know {\it a priori}
the values of $\mu$ and $\Delta$ which determine the relative error in
Eq.~(\ref{pri}), but as the Monte Carlo sampling progresses, we obtain
estimates of $\mu$ and $\Delta$ from the sample mean and standard
deviation.  As we sample further (increasing $N_{\rm mc}$), we obtain
more accurate estimates of these quantities, and can increase the
accuracy of the eigenvalue determination by increasing $m$, if needed,
so that the error is not dominated neither by the statistical Monte
Carlo nor by the systematic quantum phase estimation errors (see
App.~\ref{s:lor}).

\section*{Other applications} 
The quantum algorithm for 
estimating determinants shows the power of being able to sample from the
eigenspectrum of a matrix.   The same method of supplementing
the quantum phase estimation algorithm with Monte Carlo sampling applies to
estimating $F:=\sum_{j=1}^n f(\lambda_j)$ for any readily computable non-negative %{\bf{[PERCHE' non negativa?]}}
function $f$.   For example, if we can implement the dynamics
$e^{-iAt}$, where $A$ is the Hamiltonian for
a physical system, and $f(x) = e^{-\beta x}$, then these techniques
allow us to estimate the partition function 
$Z(\beta) = \sum_j e^{-\beta E_j}$ of the Hamiltonian.  
Similarly, if we
are given multiple copies of a density matrix $\rho$,
we can use density matrix exponentiation \cite{dme} to perform
transformations of the form $e^{-i\rho t}$ and to estimate
the entropy $S = -\sum_j p_j \log p_j$, where $p_j$ are
the eigenvalues of $\rho$.

For estimating $F$ for an arbitrary 
Eq.~\eqref{pri} is modified as
\begin{equation}
\frac{\Delta F_{\rm err}}{|F|} 
\approx \frac1{\mu_f}\Big( \frac{\sum_j |\dot f(\lambda_j)|}{2^m n} +{\Delta_f \over {\sqrt{ N_{\rm mc}}}}\Big) 
\leqslant \frac 1{\mu_f} \Big(
{|\dot f|_{max} \over 2^m}+{\Delta_f \over {\sqrt {N_{\rm mc}}}} \Big),
\labell{ddd}\;
\end{equation}
where $\mu_f = (1/n) \sum_j f(\lambda_j)$ is the average of $f$,
$\Delta^2_f$ is its variance, $\dot f = df/dx$ is the derivative of
$f$, and $|\dot f|_{max}$ is the maximum magnitude of the derivative
of $f$ over the range of eigenvalues.
\\
After the submission of this work we were made aware that a very similar algorithm appears as a subroutine for the quantum ``supervised machine learning Gaussian processes'' algorithm that was proposed in \cite{fitz}, see also \cite{luongo}.

We thank G. De Palma and S. F. E. Oliviero for useful comments and discussions.
VG and LM acknowledge financial support by MUR (Ministero
dell’Universit\'{a} e della Ricerca) through the PNRR MUR project
PE0000023-NQSTI. SL was supported by the U.S. Army Research Laboratory
and the U.S. Army Research Office under contract/grant number
W911NF2310255, and by the U.S. Department of Energy under
contract/grant number DE-SC0012704.

% \newpage\begin{centering}{\LARGE  Supplementary Material\\}\end{centering}

\appendix
\section{Notation} \label{notation} 
In this section we fix the notation and define the relevant quantities of the problem. 
Let $A$ be a (strictly) positive $n\times n$  matrix  which is $s$-sparse (i.e. in each row and column of $A$ there are no more than $s$ not-null matrix elements). 
The log-determinant of $A$ is defined as 
\begin{eqnarray}
\alpha : = \log \left({\rm det}[A]\right) = \sum_{j=1}^n \log \lambda_j\;,
\end{eqnarray}
with $\lambda_j\in \mbox{Sp}[A]$ being the $j$-th eigenvalue of $A$. 
We also introduce  the mean value and the  variance of the $\log$ spectrum of $A$,
\begin{eqnarray} 
\mu := \frac{1}{n}  \sum_{j=1}^n \log \lambda_j =\alpha/n  \;, \label{MU}  \\
\Delta^2 := \frac{1}{n}  \sum_{j=1}^n (\log \lambda_j - \mu)^2 \label{DELTA2} \;. \end{eqnarray} 
By proper rescaling  we assume the spectrum $\mbox{Sp}[A]$  to be bounded as follows 
\begin{eqnarray} 
\lambda_{\max} \geq \lambda_j \geq \lambda_{\min}\;, \qquad \forall \lambda_j \in \mbox{Sp}[A]\;,  \label{SPE}
\end{eqnarray} 
where we set 
\begin{eqnarray}\label{defLAMBDA} 
\lambda_{\max} :=1/2\;, \qquad  \lambda_{\min}:=1/(2\kappa)\;,\end{eqnarray}  the constant $\kappa\geq 1$ being an upper bound on the condition number $K(A)$ of the matrix,
i.e. 
\begin{eqnarray} \label{KAPPA} 
\kappa \geq K(A) := \frac{ \max_j \lambda_j}{\min_{j'} \lambda_{j'}}\;. 
\end{eqnarray} 
Under these conditions 
 it hence follows that $\mu$ and $\Delta$ can be bounded via $\kappa$ trough the inequalities \begin{eqnarray} 
\log 2 \leq |\mu| \leq \log (2\kappa)  
\;,  \label{UPMU} \qquad 
\Delta \leq\frac{ \log \kappa}{2}\;. 
\end{eqnarray} 

\section{Resource accounting}\labell{s:app}
In this appendix, we detail the resource accounting of the quantum
Hamiltonian simulation and of the quantum phase estimation algorithms.

The matrix $A$ can be seen as an Hamiltonian generator acting on a
quantum register $M_0$ of $n_Q=\log n$ qubits.  Assume hence the
possibility of coupling $M_0$ with a second register $M_1$ containing
$m$ qubits, via a series of control unitary operations $c-U(t)$ with
$U(t)$ the unitary gate $U(t)= \exp[i t A]$.  Under this premise
we can use the quantum phase estimation algorithm (QPA) to determine
the $j$-th eigenvalue of $A$ via the following operations:
\begin{enumerate}
\item initialize $M_0$ and $M_1$ into the quantum state $|j\rangle \otimes |O \rangle_B$ with 
$|j\rangle$ the eigenvector of ${A}$ associated with  $\lambda_j$, and $|O\rangle=|0\rangle^{\otimes m}$;
\item act on $M_1$ via $m$  Hadamard gates;
\item couple $M_1$ and $M_0$ via a sequence of $2^m$, $c-U(t)$ transformations;
\item operate on $M_1$ with the inverse Quantum Fourier Transform
  (QFT$^{-1}$) and measure $M_1$ (no measurement on $M_0$ is needed).
\end{enumerate} 
This allows one to get the value of $t \lambda_j/(2\pi)$ with an accuracy $\simeq 2^{-m}$.
In particular, by setting  $t=2\pi$, we see that $2^{-m}$ is exactly the error on the estimation  
of  $\lambda_j$.
 Considering  that the implementation of QFT$^{-1}$ on a $m$ qubit register requires  ${\cal O} (m \log m)$ logical operations, 
 the computational cost (number of logical operations required to complete the task) of the above procedure is  
\begin{eqnarray} \label{COMP1} 
N_{\rm qpe} &\simeq& 2^m  \times \#_{\rm{U}}  + {\cal O} (m \log m)  \;,
%T_{\rm qpe} &\simeq& 2^m  \times T_{\rm{U}}  + {\cal O} (m \log m) 
\end{eqnarray} 
with $\#_{\rm{U}}$  the computational cost of
implementing a single $c-U(2\pi)$ operation (of course in case we have
an oracle implementation of $c-U(2\pi)$ then we do not need to
translate the $2^m$ calls in logical gates).  The same procedure can
be applied to a uniform random sampling of the spectrum of ${\alpha}$ by
simply replacing the input state in step 1 with the density matrix
$\openone/n \otimes |O\rangle \langle O|$.  Running the algorithm $N$
times as needed by the log-determinant algorithm,  will generate a random sequence
$\tilde{\cal S} = (\lambda_{j_1}^{(1)}, \lambda_{j_2}^{(2)} ,
\cdots, \lambda_{j_{N }}^{(N)})$, uniformly distributed, each result
with an error $\delta \lambda \simeq 2^{-m}$.  The total computational
cost in this case is clearly given by $N$ times the
value~(\ref{COMP1}), i.e.
 \begin{equation} \label{COMPN} 
\#_{\rm LgD}=  N \times N_{\rm qpe}   
\simeq N 2^m \times  \#_{\rm{U}}  + {\cal O} ( N m \log m) \;.
\end{equation} 
From~\cite{berry3} we learn that in the case of an $s$-sparse
Hamiltonian ${A}$ one can implement the action of the $c-U(t=2\pi)$ with
an accuracy $\eta$ on $M_1$ by means of
\begin{eqnarray} 
\#_{\rm{query(A)}}\simeq {\cal O} \left(\tau \frac{\log(\tau/\eta)}{\log
    \log(\tau/\eta)}\right)\;,
    \end{eqnarray}  {\it queries} of the Hamiltonian ${A}$
and
\begin{eqnarray}\#_{\rm 2-q}\simeq
{\cal O} \left( \tau \log n \frac{\log^2(\tau/\eta)}{\log
    \log(\tau/\eta)}\right)\;,\end{eqnarray}  extra 2-qubit operations, where given  $\| A\|_{m}$ an upper bound on the absolute values of entries of the $n\times n$  matrix $A$, one has
\begin{eqnarray}
  \tau = 2\pi  s^2  \| A \|_{m} \;.\end{eqnarray} 
We remark that Ref.~\cite{berry3} is based on an oracle model, where it is assumed that the user has access to
a black-box  that accepts a row index $i$ and a number $j$ between $1$ and $s$, and returns the position and value of the $j$-th nonzero entry of ${A}$ in row $i$. More precisely, Ref.~\cite{berry3}   assumes the
existence of a quantum circuit that, given the indexes $i$ and $j$, gives a control-quantum phase gate 
whose value is proportional to the entry corresponding entry of ${A}$, i.e. something of the form 
 \begin{eqnarray} \label{query} 
 |i,j\rangle \otimes |b\rangle \; \; {\longrightarrow} \;\;  (-1)^{b A_{ij}t} |i,j\rangle\otimes |b\rangle \;, 
 \end{eqnarray}  
 where $|i,j\rangle$ is a quantum register that enumerates all the
 non-zero entries of ${A}$, and $|b\rangle$ with $b\in \{0,1\}$ is a
 control qubit.  Each individual operation~(\ref{query}) counts as a
 {\it query}\footnote{In a subsequent work~\cite{childs} the same
   authors of Ref.~\cite{berry3} prove that a similar scaling can be
   achieved also when $A$ is represented as sum of the form
   $\sum_{k} c_k V_k$ where $V_k$ are unitary operations which
   can be physically implemented on the register either by some oracle
   or via some quantum circuit.}; the {\it oracle complexity} of the
 algorithm~\cite{berry3} is given by $\#_{\rm{query(A)}}$ 
while 
the sum of this number  and the number $\#_{\rm 2-q}$  defines the {\it time complexity} of the algorithm, i.e.
\begin{eqnarray} \nonumber 
T^{\rm [ora]}_{\rm U}&\simeq&\#_{\rm{query(A)}} + \#_{\rm 2-q} \label{TU}\\ \nonumber 
&=& {\cal O} \left(\tau \tfrac{\log(\tau/\eta)}{\log \log(\tau/\eta)}\right)+
 {\cal O} \left(\tau  \log n  \tfrac{\log^2(\tau/\eta)}{\log \log(\tau/\eta)}\right)\nonumber \\
 &\simeq& 
 \label{timecomp} 
 {\cal O} \left(\tau  \log n  \tfrac{\log^2(\tau/\eta)}{\log \log(\tau/\eta)}\right)\;,
 \end{eqnarray} 
 where in the final  identity we dropped the smallest of the two contributions (i.e. $\#_{\rm{query(A)}}$). 
 The global computational cost  of implementing the action of  $c-U(2\pi)$ can instead be estimated as 
\begin{eqnarray}\label{DDFewerq} 
\#_{\rm{U}} &\simeq& \#_{\rm{query(A)}} \times \#_{\rm A} + \#_{\rm 2-q} \\ \nonumber 
&=&
{\cal O} \left(\tau \tfrac{\log(\tau/\eta)}{\log \log(\tau/\eta)}\right) \times \#_{\rm A}+ {\cal O} \left(  \tau  
\log n \; 
\tfrac{\log^2(\tau/\eta)}{\log \log(\tau/\eta)}\right),  
\end{eqnarray} 
where now  $\#_{\rm A}$ indicates the computational cost needed to implement a single query of  ${A}$ according to the above prescriptions, and the second contribution
represents the extra cost in terms of two-qubit gates one has to pay in order to complete the task (see also Table 1 of Ref.~\cite{REF3}).
\\ 
From Eq.~(\ref{timecomp}) it immediately follows that within the oracle model of Ref.~\cite{berry3} the time-complexity of the log-determinant algorithm can be estimated as
 \begin{eqnarray} \nonumber 
T^{\rm [ora]}_{\rm LgD} &\simeq&   N 2^m T^{\rm [ora]}_{\rm U}   + {\cal O} ( N m \log m) \nonumber \\
&\simeq&
 N 2^m  {\cal O} \left(\tau  \log n  \tfrac{\log^2(\tau/\eta)}{\log \log(\tau/\eta)}\right)   + {\cal O} ( N m \log m)\nonumber \\
&\simeq&   {\cal O} \left(  N 2^m s^2   \log n  \; {\log^2(s^2 /\eta)}\right) \;,\label{TIMEnew} 
\end{eqnarray} 
where, in the first expression, we have omitted the order-one constant factor 
 $2 \pi \| A\|_{m}$  and a subleading $\log \log$ factor, while 
 in the second line, the ${\cal O}(N m \log m)$ contribution is dropped as it is clearly dominated by the leading term.
Notice that if $\#_{\rm A}$ does not grow faster than
$\log n$, the first contribution in Eq.~(\ref{DDFewerq}) can be neglected.
In this case, $\#{\rm{U}}$ is effectively determined by the second term alone and can be identified with the time complexity cost $T_{\rm{U}}$ given in Eq.~(\ref{timecomp}):
\begin{eqnarray}
\#_{\rm{U}} &\simeq& T_{\rm{U}}  \simeq {\cal O} \left( \tau   \log n  \tfrac{\log^2(\tau/\eta)}{\log \log(\tau/\eta)}\right),
 \label{DDFewerqsds} 
\end{eqnarray} 
so that Eq.~(\ref{COMPN}) yields
\begin{equation}
\#_{\rm LgD} \simeq  T^{\rm [ora]}_{\rm LgD}  
\simeq   {\cal O} \left(  N 2^m s^2   \log n  \; {\log^2(s^2 /\eta)}\right),\label{COMPNnew} 
\end{equation} 
This condition can be met either by ensuring that the operation $A$ is implemented efficiently \cite{hast,chuang}, or by assuming that the output of a prior quantum computation has been stored in a quantum random access memory~\cite{qram,nielsenchuang}. In the latter case, all relevant performance metrics -- such as addressing time, average number of activated gates, and expected energy consumption -- scale as $\log n$, even in the worst-case scenario where all memory entries are accessed in superposition.

\section{Error budget}\labell{s:appvit2}
In this appendix, we give the details of the error budget calculated in the main text.

Consider the ideal case where the evaluation of the eigenvalues of the matrix $A$ is affected solely by the statistical error associated with Monte Carlo sampling, with no additional errors introduced by the quantum phase estimation procedure. Under this condition, after    $N$ measurements, we shall obtain a string  
 \begin{eqnarray} \label{String1} 
 {\cal S}_N:= (\lambda_{j_1}, \lambda_{j_2} , \cdots, \lambda_{j_N})\;, \end{eqnarray}
 of random elements of $\mbox{Sp}[A]$. Since the process is ruled by a uniform  probability distribution
 a good estimator of the log-determinant of $A$  is obtained by taking the average of the logarithms of the elements of  ${\cal S}_N$, 
\begin{eqnarray}  \label{EST1} 
  \alpha_{est} = \frac{n}{N} \sum_{\ell=1}^N \log \lambda_{j_\ell} =
  \frac{n}{N} \sum_{j=1}^n N_j \log \lambda_{j}\;,\end{eqnarray}
where $N_j$  is the number of times that the $j$-th
element of $\mbox{Sp}[A]$ (i.e.~$\lambda_j$) appears in ${\cal S}_N$.  
Indeed, one can easily verify that, in the limit of  large $N$, $\alpha_{est}$ approaches ${\alpha}$
with high probability,  i.e.
\begin{eqnarray} \label{asympt1} 
\alpha_{est} \Big|_{N\gg 1} \longrightarrow 
\quad \left\langle \alpha_{est} \right\rangle = \alpha\;,  
\end{eqnarray} 
 with a statistical error (square root of the associated variance)  \begin{eqnarray} 
\Delta \alpha_{\rm mc} : = \sqrt{\left\langle  
\left( \alpha_{est}  - \alpha \right)^2\right\rangle} =   n \frac{ \Delta}{\sqrt{N}}\;, \label{asympt} 
\end{eqnarray} 
with $\Delta$ defined in Eq.~(\ref{DELTA2})
(we use the symbol $\left\langle \cdots \right\rangle$ to represent
the mean with respect to the stochastic process that generates the
corresponding variable).  We recall that the operational meaning of
(\ref{asympt}) is provided by the Chebyshev inequality which states that
the probability that the distance between the estimator
$\alpha_{est}$ and ${\alpha}$ is larger than or equal to $\gamma$
can be bounded by the inequality
\begin{eqnarray} \label{errorrest10} P_{\rm rob}\left(\left| \alpha_{est} - \alpha   \right| \geq  
    \gamma\right) \leq  \frac{(\Delta \alpha_{\rm mc})^2}{{\gamma^2}}  = 
     \frac{n^2 \Delta^2}{{N\gamma^2}} \;. \end{eqnarray} 
The {\em relative} error associated with  (\ref{asympt}) is
\begin{eqnarray} 
\frac{\Delta \alpha_{\rm mc} }{|\alpha|}  =  \frac{n \Delta}{|\alpha|\sqrt{N}} = \frac{1}{|\mu|} \; \frac{ \Delta}{\sqrt{N}}  \;, 
\label{ll}
\end{eqnarray}
with $\mu$ defined in Eq.~(\ref{MU}). 
When 
 the sampling of $\mbox{Sp}[A]$ is performed with some error, the procedure will produce  
$N$ independent random outcomes which approximate the exact values
$\lambda_{j}$'s up to an error $\delta \lambda$ which is fixed by the
selected estimation procedure (e.g. the quantum phase estimation (QPA)
algorithm). In the case of QPA, this is not a statistical error, but a
systematic one: a truncation in the binary expansion. More precisely,
during the sampling, each element $\lambda_j \in \mbox{Sp}[A]$, is
replaced by an associated random variable ${\lambda}_{j}^{(k)}$
distributed according to some, for now unspecified, conditional
probability function $P(k|j)$. We shall however assume that, for
$j\in \{1,\cdots, n\}$ the random values $\lambda_{j}^{(k)}$ does not
differ from the $j$-th element of $\mbox{Sp}[A]$ more than a fixed
threshold value $\delta \lambda$, i.e.
\begin{eqnarray} \label{constraint} 
\left| {\lambda}_{j}^{(k)} - \lambda_{j}\right| <   \delta \lambda\;, \qquad \forall k\;. 
\end{eqnarray} 
To ensure the positivity of  the variables
${\lambda}_{j}^{(k)}$, in what follows we shall also assume that $\delta \lambda$ is not larger that the lower bound $\lambda_{\min}$.
In this scenario  after $N$ samplings, instead of getting the strings ${\cal S}_N$ defined in the previous paragraphs,
we actually get the noisy string 
 \begin{eqnarray}\label{String2} 
\tilde{\cal S}_N := (\lambda_{j_1}^{(k_1)}, \lambda_{j_2}^{(k_2)} , \cdots, \lambda_{j_N}^{(k_N)})\;,
 \end{eqnarray} 
 whose elements are produced with probabilities $P(j,k):= P(k|j)/n$
 ($1/n$ being the probability of getting a $j$ value from
 $\mbox{Sp}[A]$). For future purposes it is useful to establish a
 formal connection between (\ref{String1}) and (\ref{String2}):
 specifically we say that $\tilde{\cal S}_N$ is a noisy version of
 ${\cal S}_N$, if for all $\ell \in \{ 1,\cdots, N\}$, the $\ell$-th
 element of the former has the same $j$ value of the corresponding
 $\ell$-th element of the latter.

 Using the string $\tilde{\cal S}_N$ we now wish to recover ${\alpha}$ using
 a simple averaging scheme. Specifically, we assume as estimator the
 quantity
\begin{align} 
  \tilde{\alpha}_{est} := \frac{n}{N} \sum_{\ell=1}^N \log { \lambda}_{j_\ell}^{(k_\ell)}=
 \frac{n}{N}  \sum_{j=1}^n \sum_{k}  N_{j,k}   \log { \lambda}_{j}^{(k)}
\label{NEW} 
 ,\end{align}
where  $N_{j,k}$ counts the number of times the element $\lambda_j^{(k)}$ appears in the string  $\tilde{\cal S}_N$.
As in the case of (\ref{asympt1}) for large $N$, $\tilde{\alpha}_{est}$ will approaches with high probability  its
nominal expectation value, which in this case is given by
\begin{eqnarray} \label{asympt1new} 
&  \tilde{\alpha}_{est} \Big|_{N\gg 1} \to \left\langle \tilde{\alpha}_{est} \right\rangle = \tilde{\alpha}
                                       :=  n  \sum_{j=1}^n \sum_{k} P(j,k)  \log{\lambda}_{j}^{(k)}    \nonumber \\&
= 
 \sum_{j=1}^n \sum_{k} P(k|j)  \log{\lambda}_{j}^{(k)}
\;. 
\end{eqnarray} 
Of course in general $\tilde{\alpha}$ will differ from ${\alpha}$, indicating that the estimator 
$\tilde{\alpha}_{est}$ is not un-biased (not even asymptotically). 
In this case, instead of considering the variance of $\tilde{\alpha}_{est}$ we hence need to focus on the corresponding mean 
square error, i.e. the quantity
\begin{eqnarray} \label{Erro}
\Delta^2 {\alpha}_{\rm err} &:=&{ \left\langle \left(\tilde{\alpha}_{est}- {\alpha}\right)^2\right\rangle}\;, \end{eqnarray} 
which by the Chebyshev inequality bounds the error probability of the estimation via the relation
\begin{eqnarray} 
P_{\rm rob}\left(\left| \tilde{\alpha}_{est}   - \alpha   \right| \geq    \gamma\right) &\leq& \frac{(\Delta{\alpha}_{\rm err})^2 }{{\gamma^2}} \;.
\end{eqnarray} 
To evaluate  $\Delta{\alpha}_{\rm err}$  we relate it
with the genuine Monte Carlo variance $\Delta \alpha_{\rm mc}$ determined in Eq.~(\ref{asympt}) and the statistical distance
of the noise estimator (\ref{NEW}) from the
the estimator ${a}_{est}^{(N)}$ of a not-noisy string ${\cal S}_N$, i.e.
\begin{eqnarray} \label{QPEERROR}
\Delta \alpha_{\rm qpe}:=\sqrt{ \left\langle\left|\tilde{\alpha}_{est}- {\alpha}_{est}\right|^2\right\rangle}\;.\end{eqnarray} 
Specifically we write   
\begin{widetext}
\begin{eqnarray}
&&\Delta^2 {\alpha}_{\rm err} ={ \left\langle \left(\tilde{\alpha}_{est}- {\alpha}\right)^2\right\rangle}={ \left\langle \left(\tilde{\alpha}_{est}- {\alpha}_{est} + 
{\alpha}_{est} - {\alpha}\right)^2\right\rangle} \nonumber \\
&&\quad ={ \left\langle \left(\tilde{\alpha}_{est}- {\alpha}_{est}\right)^2\right\rangle + 
 \left\langle\left({\alpha}_{est} - {\alpha}\right)^2 \right\rangle+ 2  \left\langle\left(\tilde{\alpha}_{est}- {\alpha}_{est}\right)
\left({\alpha}_{est} - {\alpha}\right)\right\rangle} \nonumber\\
&&\quad \leq { \left\langle \left|\tilde{\alpha}_{est}- {\alpha}_{est}\right|^2\right\rangle + 
 \left\langle\left|{\alpha}_{est} - {\alpha}\right|^2 \right\rangle+ 2 \sqrt{ \left\langle\left|\tilde{\alpha}_{est}- {\alpha}_{est}\right|^2\right\rangle}\sqrt{ \left\langle 
\left|{\alpha}_{est} - {\alpha}\right|^2\right\rangle}} \nonumber
\\
&&\quad=\left( \sqrt{ \left\langle\left|\tilde{\alpha}_{est}- {\alpha}_{est}\right|^2\right\rangle} 
+ \sqrt{ \left\langle 
\left|{\alpha}_{est} - {\alpha}\right|^2\right\rangle}\right)^2 \qquad \Longrightarrow   \qquad
\Delta{\alpha}_{\rm err} \leq \Delta {\alpha}_{\rm qpe} + \Delta {\alpha}_{\rm mc} \;.  \label{Errore}
\end{eqnarray}
Next we use 
the 
constraint~(\ref{constraint}) to bound $\Delta {\alpha}_{\rm qpe}$.
As a first step, recalling the
definition of $\kappa$, we notice that for $j\in \{1, \cdots, n\}$
from~(\ref{constraint}) we can write
\begin{eqnarray}\label{DEF122} 
 &&  \left|\log{ \lambda}_{j}^{(k)} -\log{ \lambda_{j}}\right|  \leq \left\{ \begin{array}{lr}
  \log (\lambda_{j}+\delta \lambda) -\log (\lambda_{j})& \quad  \mbox{if ${\lambda}_{j}^{(k)} \geq \lambda_{j}$,} \\ \\
   \log (\lambda_{j}) -\log (\lambda_{j}-\delta \lambda) &  \quad \mbox{if ${\lambda}_{j}^{(k)} < \lambda_{j}$,} 
   \end{array} 
   \right.    \leq \left\{ \begin{array}{lr}
{\delta \lambda}/{ \lambda_{j}} & \qquad  \mbox{if ${\lambda}_{j}^{(k)} \geq \lambda_{j}$,} \\ \\
{\delta \lambda}/{( \lambda_{j}-\delta \lambda)} 
 &  \qquad \mbox{if ${\lambda}_{j}^{(k)} < \lambda_{j}$,} 
   \end{array} 
   \right.  \nonumber \\ 
 &&\qquad  \leq 
 \frac{ \delta \lambda}{  \lambda_{j}-\delta \lambda} \leq   \frac{ \delta \lambda}{ \lambda_{\min}-\delta \lambda} 
 =  \frac{  2 \kappa \delta \lambda}{ 1-2 \kappa \delta \lambda}  \;, 
\end{eqnarray} 
\end{widetext}
(the first inequality follows from the fact that $\log(x)$ is
monotonically increasing in $x$; the second instead follows from the
fact that $\log(x)$ is concave with first derivative $1/x$; notice
also that we explicitly made use of the fact that $\delta \lambda$ is
smaller than $\lambda_{\min}$ so that
$ \lambda_{j}-\delta \lambda \geq \lambda_{\min}-\delta \lambda > 0$).
The condition $\delta \lambda\leqslant\lambda_{min}$ can be enforced
(possibly iteratively) through a suitable choice of $m$. From the
above relation we can then establish the following inequality
\begin{eqnarray} 
&&\left| \tilde{\alpha}_{est}- {\alpha}_{est}\right|=\frac{n}{N} \left| \sum_{\ell=1}^N \log { \lambda}_{j_\ell}^{(k_\ell)}-\log { \lambda}_{j_\ell}\right|\label{NEW1}  \\\nonumber
&&\quad \leq \frac{n}{N} \sum_{\ell=1}^N\left| \log { \lambda}_{j_\ell}^{(k_\ell)}-\log { \lambda}_{j_\ell}\right|\leq \frac{  2 n \kappa \delta \lambda}{ 1-2 \kappa \delta \lambda
}, 
\end{eqnarray}
which replaced into (\ref{QPEERROR}) leads to 
\begin{eqnarray} \label{Errore1}
\Delta {\alpha}_{\rm qpe}  \leq 
 \frac{  2 n \kappa \delta \lambda}{ 1-2 \kappa \delta \lambda} 
\;.
\end{eqnarray}
From~(\ref{Errore}) and (\ref{asympt}) it hence follows that the Root Mean Square Error (RMSE) of the noisy estimator
can be bounded as
\begin{eqnarray} \label{Errore12}
\Delta{\alpha}_{\rm err} \leq   n\left(\frac{  2  \kappa \delta \lambda}{ 1-2 \kappa \delta \lambda}\ +  \frac{ \Delta}{\sqrt{N}}\right)\;. 
\end{eqnarray}
The  relative error is then bounded by
     \begin{eqnarray}  \label{DDF323}
\frac{\Delta{\alpha}_{\rm err}   }{|\alpha|} 
&\leq&     \frac{1}{|\mu|} \left(\frac{  2  \kappa \delta \lambda}{ 1- 2 \kappa \delta \lambda} +  \frac{\Delta}{\sqrt{N}}  \right)
\nonumber\\&=&
\frac{1}{|\mu|} \left(\frac{  2  \kappa 2^{-m}}{ 1- 2 \kappa 2^{-m}} +  \frac{\Delta}{\sqrt{N}}  \right)\;,\end{eqnarray}
where in the last line we set $\delta \lambda \simeq 2^{-m}$. We now impose that this error to be no
larger than a certain target value $\epsilon$ by requiring
\begin{equation} 
\frac{ \Delta  }{|\mu| \sqrt{N}} \leq \epsilon/2 \;, \qquad 
 \frac{  2  \kappa 2^{-m}}{ |\mu|(1-2 \kappa 2^{-m})}
 \leq \epsilon/2\;,
\end{equation} 
which can be satisfied e.g.  by taking 
\begin{equation}
\left\{ \begin{array}{l} 
N=N_{\rm mc}:=\left(\frac{ 2 \Delta}{|\mu| \epsilon} \right)^2, \\ \\
 2^{m} =N_{\rm qpe}:= 2 \kappa\left( \frac{ 2}{|\mu| \epsilon }+1 \right),   \end{array}\right. 
\Rightarrow \quad N  2^{m}
\simeq \frac{ 16 \kappa \Delta^2}{|\mu|^3 \epsilon^3}.
\end{equation} 
where in writing the last identity we assumed $|\mu| \epsilon \ll 1$. 
Replacing this into Eq.~(\ref{COMPNnew}) finally gives 
\begin{equation}  
\#_{\rm LgD} \simeq     {\cal O} \left(  \frac{ 16\;  \kappa\; s^2 \;  \Delta^2}{|\mu|^3 \epsilon^3}    \log n  \; {\log^2(s^2 /\eta)}\right) ,\label{COMPNnew2323} 
\end{equation} 
which (apart from irrelevant multiplicative constants) corresponds to
Eq.~\eqref{sec1} of the main text, once we equate all the errors: the
error from the sampling, the one from quantum phase estimation and the
one from the Hamiltonian simulation, namely $\epsilon=\eta$.
Equation~(\ref{COMPNnew2323}) is the computational cost required to
get the log-determinant of ${A}$ with a relative error $\epsilon$, i.e.
$\frac{\Delta{\alpha}_{\rm err}  }{|\alpha|} \leq \epsilon$ via
quantum random sampling that employs quantum phase estimation, quantum
Hamiltonian simulation and Monte Carlo sampling.

\section{Implementation of the method}\labell{s:lor}
In this appendix, we detail how the algorithm presented in the main
text can be implemented in practice.

The starting point is the choice of the relative accuracy $\epsilon$
with which we want to estimate ${\alpha}$. This, given the sparsity
$s$ and the dimension $n$ of the matrix $A$ allows us to set up the
Hamiltonian simulation algorithm. We start with seed values of
$N_{\rm mc}$ and $m$ and start to (randomly) estimate $N_{\rm mc}$
eigenvalues $\lambda_j$ using the algorithm detailed in the main text
to accuracy $\epsilon$. We then check whether the seed $m$ is
sufficiently large by recalling that it should be larger than
$\log_2[4\kappa/(\epsilon|\mu|)]$, where $\kappa$ and $\mu$ can be
estimated from the data. If $m$ is too small, we must start from
scratch with a larger $m$, which just introduces a multiplicative
constant overhead. (Alternatively, we can use the quantum phase
estimation only for the least significant digits.) Then, we repeat the
whole procedure $N_{\rm mc}$ times, with
$N_{\rm mc}\simeq(2\Delta/|\mu|\epsilon)^2$, where again $\Delta$ and
$\mu$ can be estimated from the data. Since the error
$\Delta \alpha_{\rm qpe}$ is a systematic (not statistical), the final
estimate of $|\alpha|$ will be affected by an error bar which is
skewed towards smaller values: the quantum phase estimation basically
gives a truncation of the estimated phase in the binary
representation.

\end{document}